\documentclass[prl,aps,twocolumn,floatfix,amsmath,amssymb,superscriptaddress,tightenlines]{revtex4}
\usepackage{graphicx}% Include figure files
\usepackage{epstopdf}
\newcommand{\be}{\begin{equation}}
\newcommand{\ee}{\end{equation}}
\usepackage{amsfonts}
\usepackage{bm}
\usepackage{color}
\begin{document}

\date{\today}
\title{Finite Temperature Critical Behavior of Mutual Information}
\author{Rajiv R. P. Singh}
\affiliation{Physics Department, University of California, Davis, CA, 95616}

\author{Matthew B. Hastings}
\affiliation{Duke University, Department of Physics, Durham, NC, 27708}
\affiliation{Microsoft Research, Station Q, CNSI Building, University of California, Santa Barbara, CA, 93106}

\author{Ann B. Kallin}
\affiliation{Department of Physics and Astronomy, University of Waterloo, Ontario, N2L 3G1, Canada}

\author{Roger G. Melko}
\affiliation{Department of Physics and Astronomy, University of Waterloo, Ontario, N2L 3G1, Canada}

\begin{abstract} We study mutual information for Renyi entropy of arbitrary index $n$, in interacting
quantum systems at finite-temperature critical points, using high-temperature expansion,
quantum Monte Carlo simulations and scaling theory. We find that for $n>1$, the critical behavior
is manifest at two temperatures $T_c$ and $n T_c$. For the XXZ model with Ising anisotropy,
the coefficient of the area-law has a $t\ln{t}$ singularity, whereas the subleading
correction from corners has a logarithmic divergence, with a coefficient related to
the exact results of Cardy and Peschel.  For $T<nT_c$ there is a  
constant term 
associated with broken symmetries that jumps at both $T_c$ and $n T_c$,
which can be understood in terms of a scaling function analogous to the boundary entropy of
Affleck and Ludwig.
\end{abstract}
\maketitle

The numerical study of entanglement in quantum systems,
through the entanglement entropy (EE) at zero temperature or mutual information (MI) at non-zero
temperature, 
promises to be a new approach to quantifying properties of quantum phases that cannot be detected using
traditional measures based on two-point correlation functions.  It has already been used in one
dimensional (1D) systems to identify the central charge \cite{Cardy,goldenchain}, in two dimensional (2D) systems to test the
area law in the Heisenberg model \cite{heisenberg1,heisenberg2}, and to identify a topologically
ordered spin liquid phase in a 2D spin model \cite{isakov}.

In 1D, gapless systems described by conformal field theory show logarithmic violations of the area law \cite{Cardy}.
However in 2D, the presence of an area law for a system such as the Heisenberg model implies
that the existence of gapless modes does not necessarily lead to such a violation.  Similar area-law
behavior is also observed in gapless 2D bosonic theories while non-interacting
fermions show a logarithmic violation \cite{bosons}, presumably reflecting the infinite
number of gapless modes associated with the fermi surface. The question of precisely which interacting many-body
models have enough entanglement to violate the area law is important both for identifying new phases and for
developing novel computational tools.

Even with an area law, subleading corrections to the entanglement entropy, such as those associated with corners,
can show logarithmic divergence at quantum critical points \cite{logcorner,logcorner2}.
While entanglement entropy at $T=0$ remains a key focus of current research, MI at non-zero temperature
(which reduces to EE at $T=0$) can also show universal critical behavior and has been a subject
of both theoretical and computational \cite{XXZ} studies.

From a computational point of view,
there is a clear need for new methods capable of studying 
EE or MI for large-scale quantum systems in $D>1$.
In this paper, we develop
a High Temperature Expansion (HTE) method for calculating MI for Renyi entropy of arbitrary index $n$ for lattice
models in the thermodynamic limit. Our work represents a new direction in the use of series expansions to study boundary phenomena in critical systems, 
enabling one to calculate corner exponents such as Cardy-Peschel exponents in
2D systems \cite{cardy-peschel}.

In the following, we combine HTE with quantum Monte Carlo (QMC) simulations and 
a scaling theory to obtain the critical
behavior of MI for a 2D spin-1/2 XXZ model, ${\cal H}=\sum_{\langle i,j \rangle} ( S^x_i S^x_j + S^y_i S^y_j + \Delta S^z_i S^z_j),$ with Ising anisotropy $\Delta = 4$.  
Quite generally, we find that for $n>1$, the critical behavior
manifests itself at two different temperatures $T_c$ and $nT_c$. Since this model is in the universality
class of the 2D Ising model, the singularity of the area-law term is
known to be $t\ \ln{t}$, 
where the reduced temperature $t$ is $|T-T_c|$ or $|T-nT_c|$, 
and the logarithmic divergence of the subleading corner terms can be related to the work of
Cardy and Peschel \cite{cardy-peschel}. We also find that spontaneously broken symmetries lead to a constant term in the
MI that jumps at $nT_c$ and $T_c$, described by a scaling function analogous
to the boundary entropy of Affleck and Ludwig \cite{affleck-ludwig}.

{\it Replica Calculation of MI---}
Consider a system divided into two regions $A$ and $B$, where $\rho_A$ is the reduced
density matrix on $A$.  The Renyi entropies are defined as 
\be
S_n(\rho_A)=\frac{1}{1-n} \ln \left[{ {\rm Tr}(\rho_A^n) }\right].
\ee
The von Neumann entropy $S_1$ is defined by the limit $n\rightarrow 1$.  The advantage of the $n>1$ Renyi
entropies is that they can be calculated by a ``replica method" for integer $n$ \cite{Cardy}, where
for a given inverse temperature $\beta$, one must evaluate a partition
function $Z[A,n,T]$ corresponding to a path integral on
a system with modified space-time topology.
In region $A$, the system is periodic
with period $n\beta$, while in region $B$ there are $n$ distinct sheets, each periodic with period $\beta$.  Normalizing correctly,
one has,
\be
{\rm Tr}(\rho_A^n)=\frac{Z[A,n,T]}{Z[T]^n}, \label{replica}
\ee
where $Z[T]$ denotes the partition function at temperature $T$.
This replica method was used in \cite{XXZ} for QMC simulations of $S_2$.  In this paper, we perform similar simulations (for $n=2$ and higher) and
also develop a HTE method to calculate the partition function with this modified topology in powers of $\beta$.
We are interested in determining the MI between region $A$ and its complement $B$, defined
as 
\be
I_n=S_n(\rho_A)+S_n(\rho_B)-S_n(\rho).
\ee

One important feature of this calculation is that if the given Hamiltonian has a critical point at temperature $T_c$, then
the partition function $Z[A,n,T]$ shows critical behavior at $T=nT_c$ because the path integral is periodic with period $n\beta$
in region $A$.  
If we consider a semi-infinite region $A$ (for example, dividing a 2D plane into two
half-planes) 
then $Z[A,n,T]$ will be non-analytic at $T=nT_c$.  
In contrast, the von Neumann MI, $I_1$,
should not show critical behavior at temperatures other than $T_c$.

\begin{figure} {
\includegraphics[width=1.8 in]{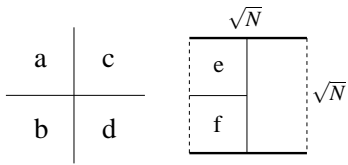} \caption{
In the HTE (left), we consider partitions of the infinite square-plane
where region $A$ could be a half-plane such as $a \cup b$ or a
quadrant such as $b$ or $c$. In QMC (right) the $N$-site real
space lattice is a torus with periodic boundaries,
or a cylinder with the dashed boundaries open.
In the torus, region $A$ can be $e$ ({``square''}) or $e \cup f$ ({``strip''}),
while for the cylindrical we use a strip ($e \cup f$) region $A$.
\label{planes}
}
} \end{figure}

{\it High Temperature Series and MI---} We develop a HTE for the MI 
in powers of $\beta=1/T$.  One important simplification of MI is that all
terms proportional to the volume of $A,B$ cancel out and we are left with only terms localized near the boundary of $A,B$.  The reason
for this is that a given bulk term in region $A$ appears once at temperature $T/n$ in $Z[A,n,T]$ but also appears once at temperature
$T/n$ in $Z[T/n]$ -- these cancel out.  Similarly, this term appears $n$ times in $Z[B,n,T]$ but also appears $n$ times in $n\ln(Z[T])$.

The HTE is calculated by a linked cluster method \cite{oitmaa,gelfand-singh-huse}. We imagine that the infinite system is divided into subregions $A$ and $B$ either by a single straight line running parallel to one of the axes, or by two perpendicular lines that meet at a point (Fig.~1). 
The line contribution to the MI is obtained by considering region $A$ to be the half-plane $a \cup b$.
To obtain the corner contribution 
we consider four separate partitions of the square 
lattice: 
the region $A$ can be (i) the quadrant $b$ (ii) the quadrant $c$ (iii) the half-plane $a \cup b$ or 
(iv) the half plane $a \cup c$.
If we add MI from the first two partitions and subtract those from the next two,
all line contributions cancel. 
The difference 
defines two times the contribution from a single corner.
More generally, we express $I_n$ as
 \be
 I_n=a_n(\beta) \cdot L+n_c b_n(\beta) +d_n(\beta), \label{linecorner}
 \ee
 where $a_n,b_n, d_n$ depend on $\beta$, $L$ is the length
of the boundary, $n_c$ is number of corners, and $d_n$ is a constant term,
associated with symmetry breaking, to be explained later.

\begin{table}
\caption{ High temperature series coefficients for the line and corner terms
for Renyi MI for $n=2$.}
\begin{tabular}{rrr}
\hline\hline
$m$  & $l_m$& $f_m$\\
2  & 1.125 & 0 \\
3  & 0.375 & 0 \\
4  & 6.32421875 & -2.765625 \\
5  &-5.109375   & 0.46875   \\  
6  & 64.02701823 &-27.11848958 \\
7  & 15.59501953 &-0.3969401042 \\
8  & 1079.586016 &-584.0700043 \\
9  & 97.15596924 &-63.38234592 \\
10 & 12847.34193 &-8700.183385 \\
11 &-1079.890682 & 94.58389488 \\
\hline\hline \label{Table1}
\end{tabular}
\end{table}

Before division by the factor $(1-n)$, the coefficient of $\beta^m$ is a polynomial in $n$ of order $m$, which vanishes at $n=0$ and $n=1$. Thus dividing by $1-n$ and taking the limit $n\to 1$ is simple and reduces the final coefficient to a polynomial of order $m-1$. 
The complete expression for the line term to $\beta^4$
is:
\begin{eqnarray}
a_n(\beta) &=& \left({\beta\over 4}\right)^2 \ {n A_2\over 2} -\left({\beta\over 4}\right)^3 \ {n(n+1) A_3\over 6} \nonumber \\
&& +\left({\beta\over 4}\right)^4 \Big[ n(n^2+n+1)\left({A_4\over 24}-{A_2^2\over 8}\right) \nonumber \\
&& +n(n^2+n-1) \left({ {B_4-A_2^2\over 2}+C_4 }\right)  \Big].
\end{eqnarray}
Here, $A_2=2+\Delta^2$, $A_3=-6\Delta$, $A_4=(2+\Delta^2)^2+4(1+2\Delta^2)$,
$B_4=\Delta^4-4\Delta^2$ and $C_4=2+\Delta^4$.
In addition, we have calculated both the line and corner contribution for the second Renyi entropy up to order $\beta^{11}$.
Let
\begin{equation}
a_2(\beta) =\sum_m l_m \beta^m, \qquad b_2(\beta)=\sum_m f_m \beta^m.
\end{equation}
The coefficients $l_m$ and $f_m$ up to $m=11$ for the second Renyi entropy 
for $\Delta=4$ are given in Table~\ref{Table1}.

{\it Comparison with Exact Numerics} --
We calculate the MI via exact diagonalization (ED), and Stochastic Series Expansion \cite{SSE} QMC using the replica-trick, Eq.~(\ref{replica}).  We extend the QMC algorithm outlined in Ref.~\cite{XXZ} to allow calculations to arbitrary Renyi entropies by directly constructing a simulation cell with $n$ sheets \footnote{We use a consistent energy normalization which does not include $n$, e.g.~in Eq.~(8) of Ref.~\cite{XXZ}.}.  Geometries considered are illustrated in Fig.~1. The critical temperature of the model is best determined by studying the Binder ratios associated with the order parameter. 
We estimate $2.234 < T_c <2.237$, which gives $\beta_c/2$ in the range $0.2235$ to $0.2238$.

\begin{figure} {
\includegraphics[width=3.2 in]{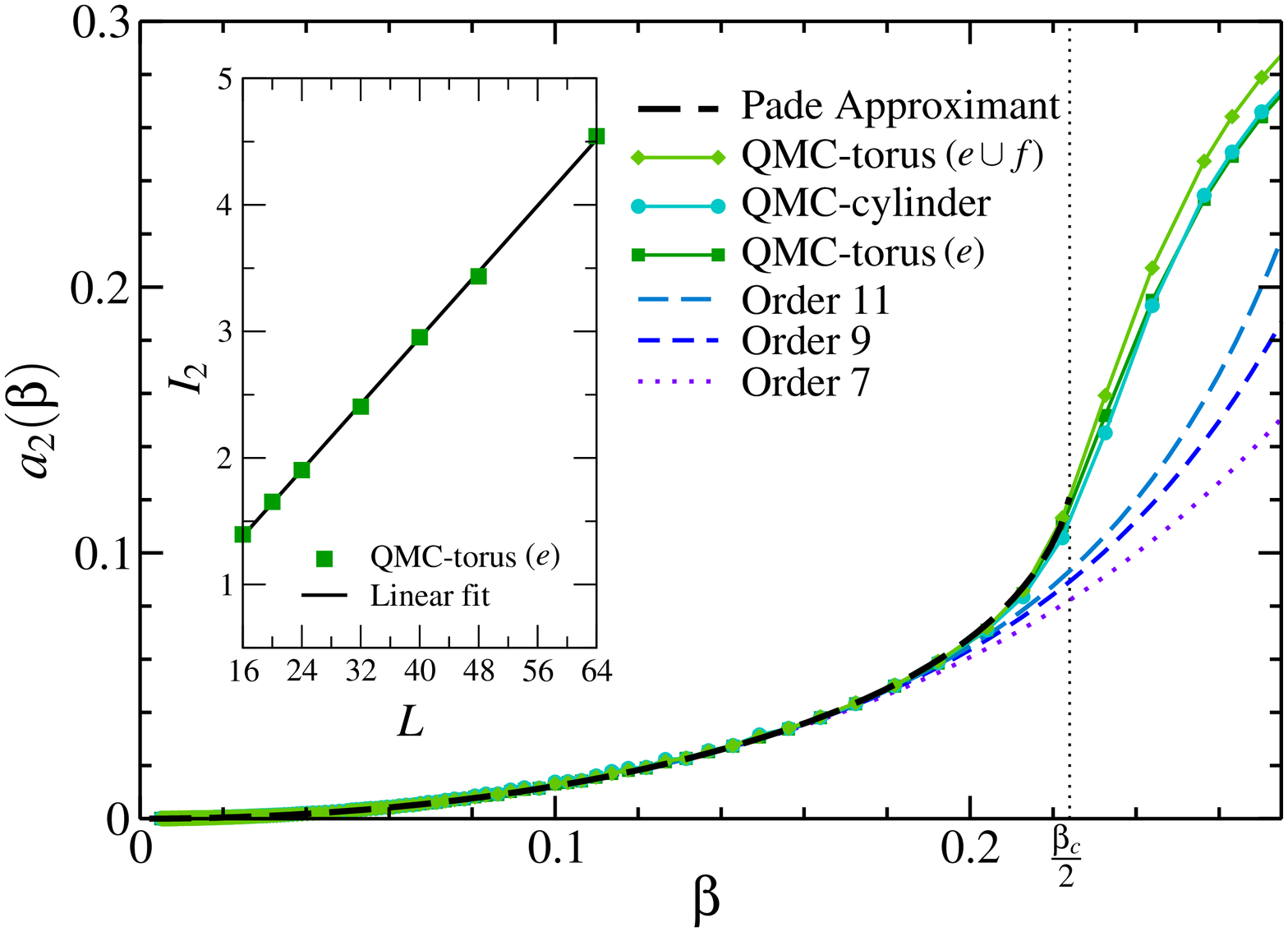} \caption{(color online) 
A comparison of MI calculated with HTE, and QMC on $32 \times 32$ simulation cells on toroid and cylindrical geometries (see Fig.~1).
Inset: MI plotted against $L$ at $\beta=0.204$, showing excellent fit to linear form.
\label{linefit}
}
} \end{figure}

Fig.~\ref{linefit}, shows the partial sums of the $11$-th order series for the linear terms (area-law)
for $n=2$ compared with 
QMC data.  One can see that the results agree extremely well up to a $\beta$ value of 
$0.21$, at which point the QMC data shows a sharp rise. To study the critical
behavior more closely we use Pade approximants. The 2D Ising universality class
is special in that the correlation length exponent $\nu=1$ and the
boundary free energy has a $t \ln{t}$ singularity \cite{au-yang-fisher}. 
Anticipating this, we take two derivatives of the series, and
use Pade approximants biased to have a pole at the $\beta_c/2$ value
obtained from the Binder ratios. Upon integration, these lead to a $t \ln{t}$
singularity. One such approximant is shown in Fig.~\ref{linefit} (thick dashed line). It
captures the sharp rise in QMC data extremely well, confirming the $t \ln{t}$ behavior 
to high accuracy.

The corner terms should have a logarithmic singularity. In fact, the series
for $db_2/ d\beta$ show good convergence for a simple pole implying
that $b_2$ goes as some constant $x$ times $ \ln{t}$. To get an accurate estimate for the
coefficient $x$, we once again bias the critical
temperature values. With the critical point biased at $0.223$ the
spread of Pade approximants leads to an estimate of $x=0.0143 \pm 0.0013$ ,
where as biasing it at $0.224$ leads to an estimate of $0.0151 \pm 0.0016$.
We can relate these coefficients to the exact results of Cardy and Peschel \cite{cardy-peschel}.
The internal angle for the corner is $\gamma=\pi/2$ for region $A$
and $\gamma= 3\pi/2$ for region B. Together with $c=1/2$ for the Ising
model, Eq.~4 in Ref.~\cite{cardy-peschel} leads to a $-{1\over 72} \ln{L}$ singularity
at the critical point. This, using $\nu=1$, translates in to an $x$ value
of ${1\over 72} =0.013\bar{8} $. Our results show that for exactly soluble
2D universality classes with known values of the central charge, the results of Cardy and Peschel can be used to obtain the coefficient $x$.

\begin{figure} {
\includegraphics[width=3.1 in]{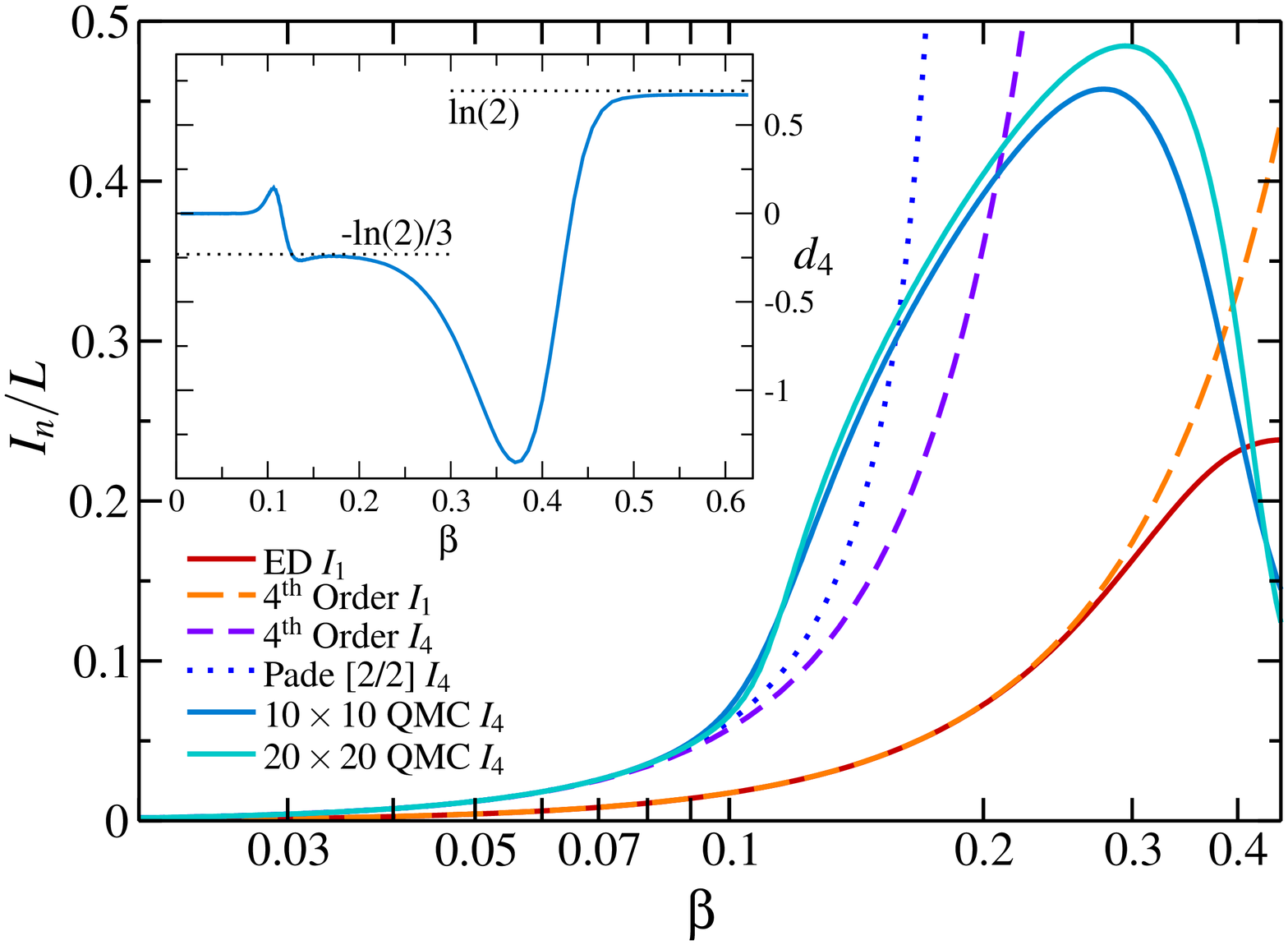} \caption{(color online) 
MI divided by boundary length from the fourth-order series expansion for: $I_1$, compared to ED on a $4 \times 4$ system with $3 \times 3$ region $A$; and $I_4$, compared to QMC for $10 \times 10$ and $20\times 20$.  A sharp change in slope in $I_4/L$ occurs at $1/4T_c$, where $T_c = 2.24$, obtained from the crossing of the fourth-order Binder cumulant for the staggered magnetization.  Inset: the constant scaling term $d_4$ extracted from a linear fit to the QMC data.
\label{fig2}
}
} \end{figure}

Fig.~\ref{fig2} shows a comparison of the von Neumann MI, calculated by continuing the HTE to $n=1$, with results obtained by exact diagonalization (ED) on a $4\times 4$ system. In this case, we have multiplied the series by the length of the boundary separating regions $A$ and $B$.
The agreement is excellent up to $\beta \approx 0.25$, which confirms the validity of both calculations and shows that finite size effects are small at smaller $\beta$ values. The von Neumann entropy series should be convergent down to $T_c$.
Fig.~\ref{fig2} also compares HTE and QMC simulation results for $I_4$, which further confirms that for $I_n$ the higher temperature singularity moves to $n T_c$.  The inset illustrates the constant scaling term $d_4(\beta)$ extracted from QMC data taken on $10 \times 10$ and $20 \times 20$ toroidal simulation cells with strip regions $A$ ($e \cup f$ in Fig.~1).  At low temperatures, $d_4(\beta)$ approaches the value $\ln(2)$ predicted from our scaling theory.  For temperatures between $T_c$ and $n T_c$, theory predicts that $d_n = -\ln(2)/(n-1)$, discussed below, which is visible as a plateau in the QMC data.

{\it Constant Terms in the MI Due to Symmetry Breaking---}
We now consider the MI between region $A$ and $B$ away from criticality, in the limit of large system size.
In addition to the line and corner
terms, symmetry breaking can lead to additional constant terms $d_n$. First, consider the case of $T<T_c$, where
the Ising symmetry is broken in all regions.  The breaking of the symmetry means that the partition functions  $Z[T],Z[T/n],Z[A,T,n],Z[B,T,n]$ all have a multiplicative factor of $2$ in addition to the volume, line, and corner terms.  The volume terms still cancel, and
the line and corner terms still contribute
according to Eq.~(\ref{linecorner}),
but the factors of $2$ increase
the MI by $\ln(2)$. 
Similarly, for $T_c<T<nT_c$, the partition function $Z(T)$ has no additional factors of $2$ but $Z(A,n,T)$, $Z(B,n,T)$ and $Z(T/n)$ do,
giving rise to a constant term in MI of $d_n=-\ln(2)/(n-1)$. These are verified by the plateau in the QMC data in the inset of Fig.~\ref{fig2}.
These results are modified strongly by finite size effects due to
the volume terms which go generically as $\exp(-L/\xi)$, but can be substantial when $\xi$ is comparable to
or larger than $L$. These form part of the scaling theory, which we develop next.

{\it Scaling Theory Near $nT_c$---} Near $T=nT_c$ we can use scaling theory to describe the singular  behavior of the
Renyi entropy.  The sheets of the system with period $\beta$ are not critical, while the region with period
$n\beta$ is in a critical scaling regime.  Consider first the case that $T>nT_c$ and $L\gg \xi$.
Then, we can calculate the MI by using a scaling ansatz for the
free energy of a critical theory with a boundary, which
implies that the singular terms in the MI
equal $c_1 (L/\xi)+c_2 n_c \ln(\xi)$ for some universal constants $c_1,c_2$.
The $c_1$ term represents the fact that the singular terms in the MI are due to degrees of freedom
at length scale $\xi$ and there are $L/\xi$ such terms. For $T<nT_c$ and $L \gg \xi$, there
is the additional $-\ln(2)/(n-1)$ described above, but the singular MI behaves again
as
$c'_1 (L/\xi)+c_2 n_c \ln(\xi)$.
For $L\sim \xi$, finite size scaling implies that
the MI is equal to $F(L t^\nu)+n_c \ln({\rm min}(L,\xi))$ plus smooth terms (such smooth terms multiplying $L$ or $n_c$) where
$F(x)\rightarrow c_1 x$ as $x\rightarrow +\infty$, $F(x)\rightarrow c'_1 |x|-\ln(2)/(n-1)$ as
$x\rightarrow -\infty$.  At $x=0$, $F(x)$ equals $n$ times the Affleck-Ludwig boundary entropy \cite{affleck-ludwig}.

The 2D Ising universality class with $\nu=1$ is special and in this case
the line term has a multiplicative log correction as verified in our
series analysis. The subleading corner term
is predicted to diverge logarithmically, in agreement with the series calculation. 
The negative jump in the
additive constant term together with an increasing line term
leads to an approximate crossing of $I_n/L$ for different system sizes 
near $nT_c$ as seen in the QMC data in Fig.~\ref{fig2}.

{\it Scaling Theory at $T_c$---} At $T$ near $T_c$, we can again develop a scaling theory.  In contrast to the
case of $T\approx n T_c$, the region with period
$n\beta$ is now in the ordered phase, and the $n$ sheets with period $\beta$ 
display critical scaling
of a theory with a boundary magnetization (since the region with period $n\beta$ is ordered).
For $L \gg \xi$, the singular terms in the MI again behave as $c_3 (L/\xi)+c_4 n_c 
\ln(\xi)-\ln(2)/(n-1)$  or $c'_3 (L/\xi) + c_4 n_c \ln(\xi)+\ln(2)$ depending on whether $T>T_c$ or $T<T_c$
with universal constants $c_3,c'_3,c_4$.
The change in sign in the constant from $-\ln(2)/(n-1)$ to $\ln(2)$
leads to a 
crossing of $I_n/L$ for different system
sizes at $T_c$ (with corrections from line and corner terms which shift
the crossings at finite $L$ to larger $T$).  There is again a multiplicative log correction in the Ising case.  This critical
point corresponds to the case of a boundary magnetic field, while the $T=nT_c$ critical point corresponds to the
case of free boundary conditions -- but both produce a log correction.

{\it Discussion---}
We have developed computational methods and scaling theory
to study Renyi mutual information $I_n$ in interacting quantum systems.  
Away from critical points the MI consists of line terms (area-law), 
corner terms, and constant terms coming from symmetry-breaking.
At the critical points the line terms develop a singularity which vanishes 
as $1/\xi$, and thus have a critical exponent $\nu$. In the special case of the 2D
Ising universality class with $\nu=1$, there are multiplicative log terms.
The subleading corner terms show a log divergence, whose coefficient can be
related to the central charge using
the results of Cardy and Peschel \cite{cardy-peschel}. We also find that
the constant terms jump discontinuously at the transitions and can be described by a scaling function that is analogous to the boundary entropy of Affleck and Ludwig \cite{affleck-ludwig}.

We have extended our previous QMC algorithm for $I_2$ \cite{XXZ} to calculate arbitrary $I_n$ by using a multi-sheeted space-time simulation cell, and confirmed the main results of the scaling theory.
QMC methods are able to access all temperature regions, allowing one to obtain
the bulk terms due to symmetry breaking.  HTE can separately obtain the line terms and subdominant corner terms.
Since the HTE is immune to the sign problem, it 
should be a general tool for calculating MI in arbitrary interacting quantum
systems such as frustrated spin or fermionic models in the future.

{\it Acknowledgments --}
We thank T. Grover and S. Isakov for useful discussions.
This work is supported by NSERC of Canada (ABK and RGM),  NSF grant No PHY 05-51164 (KITP)
and NSF grant No DMR-1004231 (RRPS). Simulations were performed using the computing facilities of SHARCNET.

%\bibliography{Biblio}

\end{document}